\documentclass[reqno,11pt]{amsart}

\usepackage{amsmath}
\usepackage{amsfonts}
\usepackage{indentfirst}
\usepackage{amssymb}

\newcommand{\bibit}[1]{\bibitem[#1]{#1}}
\newcommand{\paper}[1]{{\it #1}, }
\newcommand{\journal}[4]{#1 {\bf #2}, #3 (#4)}
\newcommand{\CMP}{Commun.\ Math.\ Phys.}
\newcommand{\HPA}{Helv.\ Phys.\ Acta}

\newcommand{\PRB}{Phys.\ Rev.\ B}
\newcommand{\PRL}{Phys.\ Rev.\ Lett.}

\numberwithin{equation}{section}
\newtheorem{theorem}{Theorem}[section]
\newtheorem{proposition}[theorem]{Proposition}
\newtheorem{lemma}[theorem]{Lemma}

\newcounter{eqs}

\renewcommand{\leq}{\;\leqslant\;}
\renewcommand{\geq}{\;\geqslant\;}

\newcommand{\upchi}{\raise1pt\hbox{$\chi$}}

\def\writefig#1 #2 #3 {\rlap{\kern #1 truecm \raise #2 truecm
\hbox{#3}}}
\def\figtext#1{\smash{\hbox{#1}} \vspace{-5mm}}
\input epsf

\begin{document}

\vspace{0.5cm}
\title{Lower Bound for the Segregation Energy in the Falicov-Kimball Model}

\author{Pedro S. Goldbaum}

\maketitle

\vspace{-0.8cm}
\begin{center}
{\small Department of Physics, Princeton University \\
Jadwin Hall, Princeton, NJ 08544, USA.}
\end{center}

\vspace{0.5cm}

\renewcommand{\thefootnote}{}
\footnote{E-mail: {\rm goldbaum@princeton.edu}.}

\begin{quote}
{\small
{\bf Abstract.}
In this work, a lower bound for the ground state energy of the 
Falicov-Kimball model for intermediate densities is derived. The explicit derivation is important in the proof of the conjecture of segregation 
of the two kinds of fermions 
in the Falicov-Kimball model, for sufficiently large interactions. This bound is given by a bulk term, plus a
 boundary term of the form $\alpha_1 (n)|\partial \Lambda|$, where $\Lambda$ is the region devoid of classical particles and $n$ is the 
density of electrons. 
A detailed proof is presented for $n=1/2$, where the coefficient $\alpha_1(1/2)=10^{-13}$ is obtained, for the two dimensional case. 
Although clearly not optimal in terms of order of magnitude, 
this is the largest explicitly calculated coefficient in the range of intermediate densities. 
With suitable modifications the method can also be used to obtain a coefficient for all densities.
 That is the topic of the last section, where a sketch of the proof for $n<1/2$ is shown. 
}

\vspace{1mm}

\end{quote}

\section{Introduction}
The Falicov-Kimball model \cite{FK} was introduced to investigate metal-insulator 
transitions in mixed valence compounds of rare earth and transition metal oxydes. Later, it was 
again considered to describe order in mixed valence systems and binary
alloys. A review of exact results for this model can be
found in \cite{GM}.

The model assumes two kinds of fermions in the lattice $\Omega$: classical (infinitely massive) `ions' with density $n_c=N_c/|\Omega|$ 
and electrons with density $n_e=N_e/|\Omega|$. For simplicity, the particles are assumed to be spinless (without loss of generality, the spin 
variable can be introduced later). 
The Falicov-Kimball hamiltonian can be written in 
the second quantized form 
\begin{equation}
H=-\sum_{x,y \in \Omega}t_{xy}a_{x}^\dagger a_{y} + U\sum_{x\in \Omega} n_x w(x),
\end{equation}
where $a_x^\dagger$ and $a_x$ are the fermion creation and annihilation operators for the electrons 
in $x$, and $n_x=a_x^\dagger a_x$. The variable $w(x)$ can be either $1$ or $0$, according to whether the site $x$ is occupied by a classical particle 
or not. We will assume here $\Omega \in {\mathbb Z}^d$

For a bipartite lattice $\Omega=A\cup B$, Kennedy and Lieb \cite{KL} proved that the ground state displays crystalline 
long range order at half filling ($n_c+n_e=1$). This result illustrates the relevance of the model in fundamental
 problems in condensed matter physics, like understanding the formation of crystals and molecules. Also, it expected that the better 
understanding of the Falicov-Kimball model will provide new insights to the Hubbard model. And in the context 
of the Hubbard model, other fundamental question can be addressed, like the existence of 
ferromagnetism in a system in which the spins are itinerant (not localized).

A long standing conjecture for the Falicov-Kimball model \cite{FF} was that, for sufficiently strong interactions, the two kinds 
of particles should segregate away from half-filling. This conjecture was proved in \cite{FLU} where it is shown 
 that the total ground state energy is bounded above and below by a bulk term, plus 
a second term which is proportional to the boundary of the region $\Lambda$ devoid of classical particles. 
If $E_{\Lambda,N}$ is the ground state energy for $N$ electrons,  
\begin{equation}
e(n)|\Lambda|+\alpha_1(n)|\partial \Lambda| \leq E_{\Lambda,N} \leq e(n)|\Lambda|+\alpha_2(n)|\partial \Lambda|, 
\end{equation}
where $e(n)$ is the energy per site for a density $n=N/|\Lambda|$ of free electrons in the infinite lattice ${\mathbb Z}^d$. 
Therefore, given that the bulk term is fixed for all configurations, 
lowering the energy requires minimizing the boundary, which is accomplished by segregating the two species of fermions 
from each other.

Also in \cite {FLU} explicit coefficients $\alpha_1(n)$ are obtained for low densities of electrons $n \leq |S_d|/(4\pi)^d$, where $\Lambda$ is the domain devoid of classical 
particles and $|S_d|$ is the volume of the d-dimensional sphere, whereas $\alpha_2(n)$ is determined for all densities. The lower bound is obtained by considering first the $U=\infty$ case. Taking $t_{xy}\equiv 1$, 
the hamiltonian acting on a function $\varphi(x)\in L^2(\Lambda)$ can be written
\begin{equation}
[h_\Lambda \varphi](x)=2d\varphi(x)-\sum_{e:x+e\in\Lambda} \varphi(x+e),
\end{equation}
where the sum is over the edges of the lattice. The eigenvalue equations are $h_\Lambda\varphi_j=e_j\varphi_j$, for $j=1,\dots,|\Lambda|$.
 Their lower bound is derived from the inequality 
\begin{equation}
E_{\Lambda,N}-|\Lambda|e(n) \geq \frac{1}{(2\pi)^d}\int (\varepsilon_F-\varepsilon_k)
\sum_{j:e_j>e_N}\frac1{(4d)^2}
|(b_k,\varphi_j)|^2 dk,
\end{equation}
where 
$\varepsilon_k=2d-2\sum_i\cos k_i$. 
Also, the concept of the boundary vector
\begin{equation}
b_k(x) = \chi_{\partial\Lambda}(x) e^{-ikx} \sum_{e: x+e \notin \Lambda} e^{-ike}
\label{defboundvec}
\end{equation}
is introduced. Therefore, the problem reduces to showing that the boundary vector has a projection in the subspace spanned by the 
largest eigenvalues. The mathematical results are bounds for the sum of the lowest eigenvalues of the Laplace operator. For the continuous Laplace operator, 
one should refer to \cite{LY}.

This will be the starting point of our study here. Our goal is to obtain an explicit coefficient 
for the boundary term for intermediate densities $|S_d|/(2\pi)^d < n < 1-|S_d|/(2\pi)^d$. We are 
going to obtain results for $U=\infty$, from which the results for finite interaction can be 
derived (see \cite{FLU}). Our main result in this limit is:
\begin{theorem}
For $d=2$ and density of electrons $n=1/2$, the ground state energy of the 
Falicov-Kimball model is bounded below by
\begin{equation}
E_{\Lambda,N}-|\Lambda|e(1/2) \geq \alpha_1(1/2)|\partial \Lambda|,
\end{equation}
where $\alpha_1(1/2)=10^{-13}$.
\end{theorem}

\section{The boundary term for $n=1/2$}
\subsection{Projection of the boundary vector}
The goal is to prove that the boundary vector $b_k$ has a projection in the subspace spanned by
the eigenfunctions $\{\varphi_j\}$, $e_j>2d$ ($h_{\Lambda}\varphi_j=e_j\varphi_j$). If we can prove that 
this projection is proportional to the boundary for a subset (of non-zero measure) of the region 
in $k$-space limited by the fermi surface of $n=1/2$ ($\varepsilon_F=2d$), the boundary term can be 
calculated.

Expanding $b_k$ in terms of the eigenfunctions of $h_{\Lambda}$ we have
\begin{eqnarray}
\nonumber
-\sum_j|(\varphi_j,b_k)|^2(e_j-2d)+2\sum_{j:e_j>2d}|(\varphi_j,b_k)|^2(e_j-2d)=
\sum_j|(\varphi_j,b_k)|^2|e_j-2d| & &\\
\nonumber
=\sum_j|(\varphi_j,b_k)|^2\frac{(e_j-2d)^2}{|e_j-2d|} \geq \frac{\|(h_\Lambda-2d)b_k\|^2}{2d}. & &
\end{eqnarray}
Therefore
\begin{equation}
\sum_{j:e_j>2d}|(\varphi_j,b_k)|^2(e_j-2d) \geq \frac{\|(h_\Lambda-2d)b_k\|^2}{4d} +
\frac{(b_k,(h_{\Lambda}-2d)b_k)}{2},
\end{equation}
and
\begin{equation}
\label{deff}
\sum_{j:e_j>2d}|(\varphi_j,b_k)|^2 \geq \frac{\|(h_\Lambda-2d)b_k\|^2}{8d^2} +
\frac{(b_k,(h_{\Lambda}-2d)b_k)}{4d}\equiv f(k).
\end{equation}

Suppose we can find $k$ such that $\varepsilon_k=2d$ and $\|(h_\Lambda-2d)b_k\|^2\geq \alpha |\partial \Lambda|$,
 for some constant $\alpha$. For such $k$, we have to consider the two possible cases:
\begin{itemize}
\item
$(b_k,(h_{\Lambda}-2d)b_k) \geq 0$
\item
$(b_k,(h_{\Lambda}-2d)b_k)<0$

\end{itemize}
We should only be concerned with the second case, where the negative contribution from the second term could cancel
out the boundary term. We claim that for $k'=k+(\pi,\pi,\dots,\pi)$, $\varepsilon_{k'}=2d$, $\|(h_\Lambda-2d)b_{k'}\|=\|(h_\Lambda-2d)b_{k}\|$ 
and $(b_{k'},(h_{\Lambda}-2d)b_{k'})=-(b_k,(h_{\Lambda}-2d)b_k)\geq 0$.

Indeed, if we consider the expansions $b_k(x)=\sum_{j=1}^{|\Lambda|}c_j\varphi_j(x)$ and $b_{k'}(x)=\sum_{j=1}^{|\Lambda|}d_j\varphi_j(x)$, and observing 
that $b_{k'}(x)=(-1)^{|x|+1}b_k(x)$ and $\varphi_{|\Lambda|-j}(x)=(-1)^{|x|}\varphi_j(x)$ we have
\begin{equation}
d_j=(\varphi_j,b_{k'})=\sum_x\varphi_j^*(x)(-1)^{|x|+1}b_k(x)=-\sum_x\varphi_{|\Lambda|-j}^*(x)b_k(x)=-c_{|\Lambda|-j}.
\end{equation}
Therefore
\begin{equation}
\|(h_\Lambda-2d)b_{k'}\|^2=\sum_j|d_j|^2(e_j-2d)^2=\sum_j|c_{|\Lambda|-j}|^2(2d-e_{|\Lambda|-j})^2=\|(h_\Lambda-2d)b_{k}\|^2,
\end{equation}
and
\begin{equation}
(b_{k'},(h_{\Lambda}-2d)b_{k'})=\sum_j|d_j|^2(e_j-2d)=\sum_j|c_{|\Lambda|-j}|^2(2d-e_{|\Lambda|-j})=-(b_k,(h_{\Lambda}-2d)b_k).
\end{equation}
where the identity $e_j+e_{|\Lambda|-j}=4d$ was used.

\subsection{A bound for $\|(h_\Lambda-2d)b_k\|^2$}
Now, it remains to show that the first term in the r.h.s of \eqref{deff} can not vanish for all $k$
in the fermi surface. First, let us consider the case d=2. Since
\begin{equation}
[(h_\Lambda-2d)b_k](x)=-\sum_e b_k(x+e)=-e^{-ik\cdot x}\sum_{e:x+e\in \partial \Lambda}\sum_{x+e+e' \notin \Lambda}e^{-ik\cdot (e+e')},
\end{equation}
the absolute value will be given by a sum of exponentials over some of the second nearest neighbors $x+e+e'$. The 
following diagrams illustrate the real part of the terms associated with each site, for particular values of $k$. 
Note that the configuration defines which terms will be in the sum. If
we can prove, for suitable values of $k$, that all the terms have
positive (or negative) real part, we conclude that they are not
canceled out by each other, and a lower bound is obtained.

\setlength{\unitlength}{0.8cm}
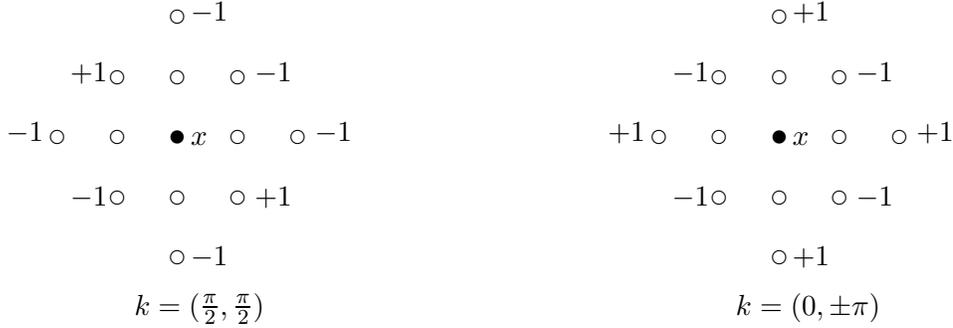
\begin{figure}[h]
\label{diagram}
\centering
\begin{picture}(0,5.5)

\put(-5,5){\circle{0.2}}
\put(-5,4){\circle{0.2}}
\put(-6,4){\circle{0.2}}
\put(-4,4){\circle{0.2}}
\put(-7,3){\circle{0.2}}
\put(-6,3){\circle{0.2}}
\put(-5,3){\circle*{0.2}}
\put(-4,3){\circle{0.2}}
\put(-3,3){\circle{0.2}}
\put(-6,2){\circle{0.2}}
\put(-5,2){\circle{0.2}}
\put(-4,2){\circle{0.2}}
\put(-5,1){\circle{0.2}}

\put(5,5){\circle{0.2}}
\put(5,4){\circle{0.2}}
\put(6,4){\circle{0.2}}
\put(4,4){\circle{0.2}}
\put(7,3){\circle{0.2}}
\put(6,3){\circle{0.2}}
\put(5,3){\circle*{0.2}}
\put(4,3){\circle{0.2}}
\put(3,3){\circle{0.2}}
\put(6,2){\circle{0.2}}
\put(5,2){\circle{0.2}}
\put(4,2){\circle{0.2}}
\put(5,1){\circle{0.2}}
\end{picture}

\figtext{
\writefig 3.55 2.75 {$x$}
\writefig 3.55 4.4 {$-1$}
\writefig 3.55 1.15 {$-1$}
\writefig 4.4 3.6 {$-1$}
\writefig 4.4 1.95 {$+1$}
\writefig 1.95 3.6 {$+1$}
\writefig 1.95 1.95 {$-1$}
\writefig 5.2 2.8 {$-1$}
\writefig 1.1 2.8 {$-1$}
\writefig 2.8 0.5 {$k=(\frac\pi2,\frac\pi2)$}
\writefig 11.55 2.75 {$x$}
\writefig 11.55 4.4 {$+1$}
\writefig 11.55 1.15 {$+1$}
\writefig 12.4 3.6 {$-1$}
\writefig 12.4 1.95 {$-1$}
\writefig 9.95 3.6 {$-1$}
\writefig 9.95 1.95 {$-1$}
\writefig 13.2 2.8 {$+1$}
\writefig 9.1 2.8 {$+1$}
\writefig 10.8 0.5 {$k=(0,\pm\pi)$}
}
\caption{Diagram of second nearest neighbors.}
\end{figure}

For $x \in \Lambda$, let $[Q_x]_{ij}=q_{x,ij}=\#\{(e,e'):e \parallel
i,e' \parallel j, x+e\in \partial \Lambda, x+e+e' \notin \Lambda \}$. 
If $tr Q_x \neq 0$
\begin{equation}
\frac{1}{2^d}\sum_{i=1}^{2^d}|[(h_\Lambda-2d)b_{k_i}](x)|^2 \geq \Bigl| \frac{1}{2^d}\sum_{i=1}^{2^d}[(h_\Lambda-2d)b_{k_i}](x) \Bigr|^2 \geq 1,
\end{equation}
where the sum is take over $k_i\in \{(\pm\frac{\pi}{2}, \pm\frac{\pi}{2}),(\pm\frac{\pi}{2}, \mp\frac{\pi}{2})\}$. Therefore we can conclude
\begin{equation}
\|(h_\Lambda -2d)b_{k_i} \|^2 \geq \#\{x \in \Lambda, trQ_x \neq 0 \},
\label{caseI}
\end{equation}
for $k_i=(\pm\frac{\pi}{2}, \pm\frac{\pi}{2})$ or $(\pm\frac{\pi}{2},
\mp\frac{\pi}{2})$. The same kind of argument makes \eqref{caseI}
valid for $d=3$ and $k_i=(\pm\frac{\pi}{2}, \pm\frac{\pi}{2},
\pm\frac{\pi}{2})$ or some vector obtained by inversion of coordinates.
 
On the other hand, if $trQ_x=0$ and $Q_x \neq 0$, $|(h_\Lambda-2d)b_k(x)|^2 \geq 1$ and
\begin{equation}
\|(h_\Lambda-2d)b_k\|^2 \geq \#\{x \in \Lambda: trQ_x=0 \ and \ Q_x\neq 0\},
\end{equation}
for $k=(0,\pm \pi)$. For $d=3$, the analogous result would be
\begin{equation}
\|(h_\Lambda-2d)b_k\|^2 \geq \frac13\#\{x \in \Lambda: trQ_x=0 \ and \ Q_x\neq 0\},
\label{caseII3d}
\end{equation}
for $k_i=(0,\frac{\pi}{2},\pi)$ or some vector obtained by permutation
of the coordinates. 

If there are no isolated sites in $\Lambda$,
\begin{equation}
\#\{x \in \Lambda: Q_x \neq 0\}=\alpha|\partial \Lambda| \ \ , \ \ \alpha \geq \frac{1}{2d}.
\end{equation}

The reason that we need not consider the case where some of the sites in $\Lambda$ are isolated 
lies in the fact that there is always a configuration obtained by joining this site to a 
larger cluster, preserving the boundary. We only need to show that the energy of the new 
configuration ($\Lambda'$) is lower than the original one.

If we have a cluster and a disjoint site, the hamiltonian can be written as a direct sum
\begin{equation}
h_\Lambda=
\begin{pmatrix}
h_1 & 0^T \\
0 & 2d
\end{pmatrix},
\end{equation}
where $h_1$ is the hamiltonian for the cluster. Consider now the perturbed hamiltonian
\begin{equation}
h(\lambda)=
\begin{pmatrix}
h_1 & v(\lambda)^T \\
v(\lambda) & 2d
\end{pmatrix},
\end{equation}
where $v(\lambda)=(0,\dots,0,-\lambda,0,\dots,0)$, such that $h_{\Lambda}=h(0)$ and 
$h_{\Lambda'}=h(1)$. We know that the sum of first $N$ eigenvalues ($E_N$) is a concave function 
of the perturbation $\lambda$. Also, there is a unitary transformation that takes 
$\lambda \rightarrow -\lambda$, which implies that each eigenvalue is an even function of 
$\lambda$. Combining these two results, we see that the sum of the eigenvalues is a 
decreasing function of $\lambda$, and $E_{\Lambda,N} \geq E_{\Lambda',N}$. So, from this point on
 we can consider $\Lambda$ as a single cluster.

For simplicity, the remainder of the proof will be
presented for $d=2$, but the method is clearly general for arbitrary
$d$. The only difference lies in the choice of the vectors $k_i$. We are going to consider the two following possible cases:
\begin{itemize}
\item
Case I: $\#\{x \in \Lambda: \ trQ_x\neq 0\} \geq \frac{\alpha}{2}|\partial \Lambda|$

\vspace{1mm}

$\|(h_\Lambda -2d)b_{k_i} \|^2 \geq \frac{\alpha}{2}|\partial \Lambda|$ for 
$k_i=(\pm\frac{\pi}{2}, \pm\frac{\pi}{2})$ or $(\pm\frac{\pi}{2}, \mp\frac{\pi}{2})$.  

\vspace{1mm}

\item
Case II: $\#\{x \in \Lambda: \ Q_x\neq 0 \ and \ trQ_x=0\} \geq \frac{\alpha}{2}|\partial \Lambda|$

\vspace{1mm}

$\|(h_\Lambda -2d)b_{k_i} \|^2 \geq \frac{\alpha}{2}|\partial \Lambda|$ for 
$k_i=(0, \pm \pi)$.  
\end{itemize}

For each of the two cases we have
\begin{equation}
f(k_i)\geq \frac{\alpha}{2^4d^2}|\partial \Lambda|.
\end{equation}
Now we need to know how rapidly can $f(k)$ vary.
\subsection{A bound for $|\nabla_j f(k)|$}
\begin{lemma}
For $f(k)$ defined by \eqref{deff}, the $j$-component of the gradient
is bounded by 
\begin{equation}
\Bigl|\nabla_j\frac{f(k)}{|\partial \Lambda|}\Bigr| \leq 10 \alpha d^3.
\end{equation}
\end{lemma}
{\bf{Proof}}: First, we should write
\begin{equation}
\frac{\partial}{\partial
k_j}|(h_{\Lambda}-2d)b_k(x)|^2=-i\sum_{e_1,e_2,e_3,e_4}(e_1+e_2-e_3-e_4)_je^{-ik(e_1+e_2-e_3-e_4)},
\end{equation}
where the sum is taken over the edges such that $x+e_1,x+e_2\in \partial
\Lambda$ and $x+e_1+e_3,x+e_2+e_4 \notin \Lambda$. We can bound the expression
in parenthesis by $4$, and the number of terms by $(2d)^4$. Also, the
number of sites where $|(h_{\Lambda}-2d)b_k(x)|^2$ does not vanish is
limited by $\alpha |\partial \Lambda|$. Therefore
\begin{equation}
\label{lemmabound1}
\big|\frac{\partial}{\partial k_j}\|(h_{\Lambda}-2d)b_k\|^2\bigr| \leq \alpha
2^6d^4|\partial \Lambda|.
\end{equation}
The same kind of argument leads to 
\begin{equation}
\label{lemmabound2}
\bigl|\frac{\partial}{\partial k_j}(b_k,(h_\Lambda-2d)b_k)\bigr|\leq 6\alpha
d^3|\partial \Lambda|,
\end{equation}
and combining the two results we conclude the proof of the lemma.

So, for 
\begin{equation}
\label{vicinity}
|k-k_i| \leq \frac{1}{10\alpha d^3}\cdot\frac{\alpha}{2^5d^2}=\frac{1}{10\cdot2^4d^5},
\end{equation}
we have
\begin{equation}
\frac{f(k)}{|\partial \Lambda|}\geq \frac{\alpha}{2^5d^2} \geq \frac{1}{2^6d^3}.
\label{bound}
\end{equation}

The lemma used to bound the gradient of $f(k)$ is useful in determining a result for any dimension. 
However, if we focus on determining a better coefficient for $d=2$, for instance, we should improve 
inequality \eqref{lemmabound1}. Instead of using the bound $(e_1+e_2-e_3-e_4)_j\leq 4$, 
we can sum over all possible vectors, using the real value of the expression in parenthesis. The same can 
be done for \eqref{lemmabound2}. The lower bound obtained for the $j$-component of the gradient of $f(k)$ is $7\alpha|\partial \Lambda|$. 
Therefore, it turns out that \eqref{bound} is valid for the extended region
\begin{equation}
\label{vicinity2}
|k-k_i| \leq \frac{1}{7\alpha}\cdot\frac{\alpha}{2^5d^2}=\frac{1}{7\cdot 2^7}.
\end{equation}

We should make a remark concerning the fact that we don't know in
principle which value of $k_i$ is the right one for Case I. But since
$\varepsilon_k$ is invariant under inversion of coordinates, the
result will be the same, regarless of the choice between the
neighborhood around $k_i=(\pm \frac{\pi}{2},\pm \frac{\pi}{2})$ or $k_i=(\pm \frac{\pi}{2},\mp \frac{\pi}{2})$ 

\subsection{The lower bound}

We are ready now to calculate the boundary term on the lower bound for the ground state energy.
First, we should recall that
\begin{equation}
E_{\Lambda,N}-|\Lambda|e(1/2) \geq \frac{1}{(2\pi)^d}\int (2d-\varepsilon_k)
\sum_{j:e_j>2d}\frac1{(4d)^2}
|(b_k,\varphi_j)|^2 dk.
\end{equation}
But we proved that 
\begin{equation}
\sum_{j:e_j>2d}|(b_k,\varphi_j)|^2\geq f(k)|\partial \Lambda | \geq \frac{1}{2^{6}d^3}|\partial \Lambda | 
\end{equation}
in the neighborhood of $k_i$. We are ready now to state the preliminary result
for $d=2$.

\begin{proposition}
Let region I be the neighborhood of $k_i=(
\pi/2, \pi/2)$ defined by $\varepsilon_k<\varepsilon_F(1/2)$ and
\eqref{vicinity2}, and region II be the neighborhood of $k_i=(0,\pm \pi)$
defined in a similar way. A lower bound for the ground state energy at $n=1/2$ is given by
\begin{equation}
E_{\Lambda,N}-|\Lambda|e(1/2) \geq \frac{|\partial \Lambda|}{(2\pi)^d}\cdot 
\frac{1}{2^8d^4} \min_{I,II}\int_{I,II} (2d-\varepsilon_k) dk=\alpha_1(1/2)|\partial \Lambda|.
\end{equation}
where $\alpha_1(1/2)>10^{-17}$.
%Explain the two factors of two. states are taken into the image states!
\end{proposition}
A similar result holds for $d=3$. The regions I and II will be defined
as the vicinities of the vectors $k_i$ presented in the last
section. Also, we should include a factor of $1/3$, to take into
account \eqref{caseII3d}.

\section{d=2: A better result}

Considering a diagram like in figure 1 for the vector $k_i=(k,\pi-k)$,
 we see that if $Q_x\neq 0$, $trQ_x=0$ and $\cos{2k}>0$,
\begin{equation}
|(h_\Lambda-2d)b_{k_i}(x)|\geq \cos{2k},
\end{equation}
whereas if $trQ_x \neq 0$ and $\cos{2k}<0$
\begin{equation}
\sum_{k_i}|(h_\Lambda-2d)b_{k_i}(x)|\geq -4\cos{2k},
\end{equation}
where the sum is taken over $k_i \in \{(k,\pi-k),(-k,\pi-k),(k,-\pi+k),(-k,-\pi+k)\}$.
Therefore we can extend the region of integration, as shown in the figure below.
The shape of the internal boundary curve is defined by the $\cos^2{(2k)}$ dependence. The 
new lower bound will be given by
\begin{equation}
E_{\Lambda,N}-|\Lambda|e(1/2) \geq \frac{|\partial \Lambda|}{(2\pi)^d}\cdot 
\frac{1}{2^9d^4} \min_{I,II}\int_{I,II} (2d-\varepsilon_k) \cos^2{(2k_x)} dk=\alpha_1(1/2)|\partial \Lambda|.
\end{equation}
Calculating the integral we get $\alpha_1(1/2)>10^{-13}$, which proves our main result.

\begin{figure}[htb]
\begin{center}
\leavevmode
\hbox{%
\epsfxsize=3.0in
\epsffile{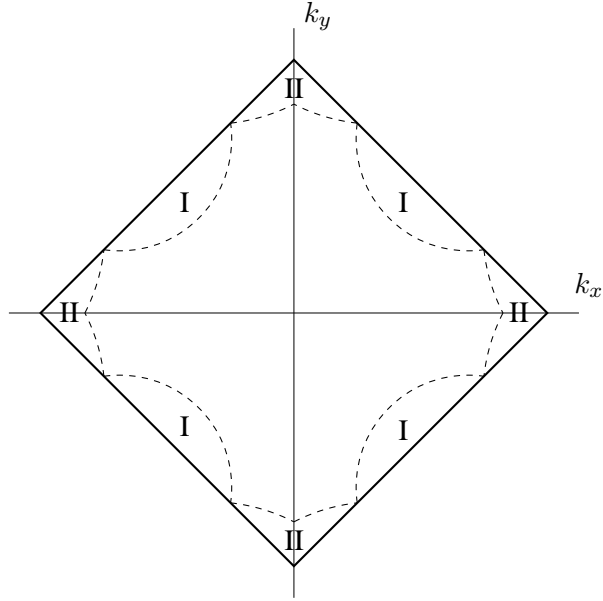}}
\caption{Fermi surface $\varepsilon_k=4$ and extended regions of integration.}
\label{fig}
\end{center}

\figtext{
\writefig 11.1 5.5 {$k_x$}
\writefig 7.5 9.1 {$k_y$}
}
\end{figure}

\section{The result for $n<1/2$}
For simplicity, we presented the detailed proof for $n=1/2$. We should stress, however, that the method is quite general,
 and can be used to obtain the lower bound for the boundary term for any density $n$. 
Taking $n=N/|\Lambda|$, we have an inequality which is equivalent to \eqref{deff}:
\begin{equation}
\sum_{j:e_j>e_N}|(\varphi_j,b_k)|^2 \geq \frac{\|(h_\Lambda-e_N)b_k\|^2}{8d^2} +
\frac{(b_k,(h_{\Lambda}-e_N)b_k)}{4d}.
\end{equation}

Again, if $(b_k,(h_{\Lambda}-e_N)b_k)<0$, we take the application $k'=k+(\pi,\pi,\dots,\pi)$, 
and we have $\varepsilon_{k'}=4d-\varepsilon_k$, $\|(h_{\Lambda}-e_N)b_k\|^2=\|(h_{\Lambda}-e_{|\Lambda|-N})b_{k'}\|^2$ 
and $(b_{k'},(h_{\Lambda}-e_{|\Lambda|-N})b_{k'})=-(b_k,(h_{\Lambda}-e_N)b_k)$.

Suppose we can prove $\|(h_{\Lambda}-e_N)b_k\|^2 \geq \alpha |\partial \Lambda|$ for some $k$ in the fermi surface. 
Then, either $\sum_{j:e_j>e_N}|(\varphi_j,b_k)|^2 \geq \alpha'|\partial \Lambda|$, or $\sum_{j:e_j>e_{|\Lambda|-N}}|(\varphi_j,b_{k'})|^2 \geq \alpha ' | \partial \Lambda|$, 
which means that the boundary contribution can be calculated for density $n$ or $1-n$. But due to particle-hole symmetry, the boundary term should be the same for 
the two densities. Therefore, the problem reduces to proving $\|(h_{\Lambda}-e_N)b_k\|^2 \geq \alpha |\partial \Lambda|$.

This observation, when combined with the lower bound for the minimum of $\|(h_{\Lambda}-e_N)b_k\|^2$ over the fermi surface 
obtained in \cite{FLU} is enough to determine the lower bound for the boundary term. To obtain a better and explicit coefficient, however, one 
should proceed like in the last sections, find vectors $k_i$ such that the function $f(k)$ cannot vanish for all of them in arbitrary configurations, 
and take the integrals over the neighborhoods of these points in $k$-space. The difference know is that we also need to perform the integration 
over neighborhoods of the vectors $k_i'$, situated in the image fermi surfaces of density $1-n$. Then we take the minimum over all of these 
integrals, to determine the lower bound.
Therefore, the choice of the vectors $k_i$ might depend on the density, but apart from that, the method is quite general.    

\section{Conclusions}
We showed here how to derive the lower bound for the boundary term of the ground state energy of Falicov-Kimball model. 
The existence of the boundary term is important since the system will try to minimize the boundary (to some extent) in 
order to minimize energy. Therefore, a segregated phase, where electrons and classical particles try to occupy distinct regions of the 
lattice, is obtained. When contrasted to the half-filling case, where crystalline long range order is observed, it might mean that the model has a first 
order phase transition when varying the chemical potentials.

Our coefficient for intermediate densities is small when compared to the upper bound obtained in \cite{FLU}. This means that our energy is
not very sensitive with respect to the boundary size. However, the strength of the method is that it provides an explicit coefficient. Also, 
$\alpha_1(n)$ can indeed be much smaller than $\alpha_2(n)$, since the upper bound is saturated by configurations with isolated sites, whereas 
the lower bound is not.

The author is indebted to E.\ H.\ Lieb and D.\ Ueltschi for helpful comments and suggestions.

\end{document}